\documentclass[twocolumn,floats,floatfix,nature]{revtex4}

\usepackage{amsfonts}
\usepackage{amssymb}
\usepackage{amsmath}
\usepackage{calc}
\usepackage[dvips]{graphicx}
\usepackage{bm}
\usepackage{mathrsfs}
\def\Black{}\def\black{\Black}

\def\be{\begin{equation}}
\def\ee{\end{equation}}
\def\bea{\begin{eqnarray}}
\def\eea{\end{eqnarray}}
\def\bse{\begin{subequations}}
\def\ese{\end{subequations}}

\def\1{\mathbf{1}}

\begin{document}

\author{Lachezar S. Simeonov}
\affiliation{Department of Physics, Sofia University, James Bourchier 5 blvd, 1164 Sofia, Bulgaria}

\title{Mechanical model of Maxwell's equations and of Lorentz transformations}

\begin{abstract}

We present a mechanical model of a quasi-elastic body (aether) which reproduces Maxwell's equations with charges and currents. Major criticism \cite{Sommerfeld1964} against mechanical models of electrodynamics is that any presence of charges in the known models appears to violate the continuity equation of the aether and it remains a mystery as to where the aether goes and whence it comes. We propose a solution to the mystery - in the present model the aether is always conserved. Interestingly it turns out that the charge velocity coincides with the aether velocity. In other words, the charges appear to be part of the aether itself. We interpret the electric field as the flux of the aether and the magnetic field as the torque per unit volume. In addition we show that the model is consistent with the theory of relativity, provided that we use Lorentz-Poincare interpretation (LPI) of relativity theory. We make a statistical-mechanical interpretation of the Lorentz transformations. It turns out that the length of a body is contracted by the electromagnetic field which the molecules of this same body produce. This self-interaction causes also delay of all the processes and clock-dilation results. We prove this by investigating the probability distribution for a gas of self-interacting particles. We can easily extend this analysis even to elementary particles.

\end{abstract}

\pacs{
}

\maketitle

\section{Introduction}
Contrary to customary views, a special reference frame and superluminal velocities are quite consistent with the theory of relativity. There are three very different empirically equivalent interpretations of relativity theory \cite{Craig2001} and one of them - LPI can quite easily accommodate Lorentz transformations with a single special reference frame and superluminal velocities (for more details on the various interpretations of relativity theory see the Appendices). LPI simply means that the physical clocks and rods have been distorted by the force fields, and Lorentz transformations connect reference frames which measure space and time with such distorted instruments. According to Bell \cite{Bell1986} this is the 'cheapest solution' in order to reconcile EPR experiments and relativity theory. Indeed, Ives \cite{Ives1945}, Builder \cite{Builder1958} and Prokhovnik \cite{Prokhovnik1993} have developed LPI and reduced it to as few a number of postulates as the familiar relativistic and Minkowskian interpretations. In this manner LPI has become as much elegant and simple as the other two interpretations. LPI does not use an arbitrary convention of the simultaneity of distant events \cite{Reichenbach1957} (putting the famous $\epsilon=1/2$) and is not based on defunct positivistic principles \cite{Craig2001}. In addition it does not unite time and space (as in Minkowskian interpretation) and treats space-time diagrams at the level of pressure-volume diagrams, i.e. instrumentally not realistically \cite{Craig2001}. On the other hand the presence of a single special reference frame is well grounded from the observational point of view \cite{Planck2019}. Indeed, it is well known that if the universe is homogeneous and isotropic, as assumed by the standard $\Lambda$CDM cosmological model, there exists a special reference frame in rest with the average motion of the cosmic matter. However, our universe is in fact statistically isotropic \cite{Weinberg2008} to one part in $10^{5}$ as can be seen\cite{Planck2019} from the statistical isotropy of the CMB. The presence of a special reference frame shows that we could attempt a mechanical explanation of the physical fields using an aether. According to Whittaker \cite{Whittaker1960} the best candidate for the aether, as some substance  with properties is the quantum vacuum. Maudlin\cite{Maudlin2018} gives clear criteria necessary for the ontology of any physical theory and a mechanical picture of electrodynamics is one such clear ontology. Perhaps the most famous and successful model of the electromagnetic field is that of MacCullagh \cite{Whittaker1960}. MacCullagh considers a continuous medium with anti-symmetric stress tensor, as shown in Fig. \ref{fig1}. The strain-stress relations in his model are,
\begin{equation}
\sigma_{ik}=\rho c^{2}\left(\partial_{i}u_{k}-\partial_{k}u_{i}\right),\notag
\end{equation}
where $\sigma_{ik}$ is the stress-tensor, $\textbf{u}$ is the displacement vector of the aether and $\rho$ is the aether's density. As can be seen from Fig.\ref{fig1} the above strain-stress relations lead to torque $d\textbf{M}=2\rho c^{2}dV\nabla\times \textbf{u}$ acting on the volume $dV$. In other words the aether resists rotations, rather than distortions, since the torque $d\textbf{M}$ is proportional to the angle of rotation $\frac{1}{2}\nabla\times \textbf{u}$. This is similar to Hooke's law for elastic media, where the force is proportional to the displacement. The equations of motion $\rho \dot{v}_{i}=\partial_{k}\sigma_{ki}$ as well as the substitution $\textbf{E}=\textbf{v}$, $\textbf{B}=-c\nabla\times \textbf{u}$ leads immediately to the familiar Maxwell's equations in vacuum. The equation $\nabla\cdot \textbf{E}=0$, which is equivalent to $\nabla\cdot \textbf{v}=0$ assumes that the aether is incompressible.

The equations MacCullagh produced are equivalent to Maxwell's equations without charges. Lord Kelvin proposed a model with symmetric stress tensor, which however is equivalent to MacCullagh's theory \cite{Whittaker1960}.

Larmor introduces charge density $\varrho$ in the MacCullagh's model \cite{Larmor1900} by simply postulating $\nabla\cdot\textbf{E}=\varrho$. If in the Larmor's model the electric field is interpreted as the velocity of the aether $\textbf{v}$, the continuity equation of the aether is obviously violated.  Indeed, if $\textbf{E}=\textbf{v}$ then in the presence of charges $\nabla\cdot \textbf{v}\neq 0$. This however contradicts the other assumption $\rho=\text{const}.$ In the model we present here, unlike Larmor's model, we do not assume an aether with a constant density, we explain the charge density $\varrho$ and at the same time, the aether is always conserved. Interestingly, it turns out that the velocity of any charge at any point in the aether coincides with the aethereal velocity at that point, which leads us to conclude that the charges are part of the aether. 

The paper is organized as follows. In section II we give a mechanical interpretation of the charges, the electric and magnetic field. In Section III we show that the model is consistent with the theory of special relativity by providing a mechanical interpretation of the Lorentz transformations. In Section IV we give the conclusions and show a path toward a mechanical picture of Einstein's gravity theory by presenting a mechanical model of the linearized Einstein's gravity equations. In the Appendices we explain fully the three interpretations of relativity theory.

\section{The model}
We examine a quasi-elastic continuous body, by which we mean a body with anti-symmetric stress tensor,
\begin{equation}
\sigma_{ik}=c^{2}\left(\partial_{i}A_{k}-\partial_{k}A_{i}\right),\;\;i,k=x,y,z\label{stresstensor}
\end{equation}
where $A_{i}$ are the components of a vector field $\textbf{A}(\textbf{r},t)$ defined by the equation:
\begin{equation}
\dot{\textbf{A}}=\rho\textbf{v}+\nabla\phi\label{model1}.
\end{equation}
Here $\rho$ is the aether density, $\textbf{v}$ is the aether local velocity field and $\phi$ is a scalar to be considered later. However note that adding $\nabla\phi $ in Eq. \eqref{model1} to $\textbf{A}$ does not alter the stress tensor. 

From Newton's equations $\rho\dot{v}_{i}=\partial_{k}\sigma_{ki}$, we have,
\begin{equation}
\rho\dot{\textbf{v}}=-c^{2}\nabla\times\left(\nabla \times\textbf{A}\right). \label{model2}
\end{equation}
Next, we include the continuity equation
\begin{equation}
\dot{\rho}+\nabla\cdot\left(\rho\textbf{v}\right)=0.\label{model3}
\end{equation}
 Equations (\ref{stresstensor}-\ref{model3}) are the basic axioms of our model.

The charge $Q$ contained in a volume $V$ is defined as:
\begin{equation}
Q\equiv-\frac{\partial}{\partial t}\int_{V}\rho dV, \label{Charge}
\end{equation}
where we have integrated along the volume $V$ (the charge is not necessarily stationary). It is obvious that if $Q>0$, aether is blasted away from the volume $V$ and if $Q<0$ aether is drawn in, towards the charge. Thus, a positive charge resembles a fan, which blasts away air but does \textit{not} produce it. Eq. \eqref{Charge} introduces a direction in time, which depends on the sign of the charge. This ought to be the case in any mechanical picture of the charge, since the electric field 'leaves' the positive charges and 'approaches' the negative charges. However if the total charge in the universe is 0, there is no global violation of time symmetry . If we examine an infinitesimal volume equation \eqref{Charge} implies
\begin{equation}
\varrho=-\dot{\rho}.\label{chargedensity}
\end{equation}
The electric field $\textbf{E}(\textbf{r},t)$ is defined as the flux of the aether, i.e.
\begin{equation}
\textbf{E}\equiv\rho \textbf{v},\label{electricfield}
\end{equation}
while the magnetic field $\textbf{B}(\textbf{r},t)$ is defined as
\begin{equation}
\textbf{B}\equiv-c\nabla\times\textbf{A}.\label{magnetic field}
\end{equation}

\begin{figure}[tb]
\includegraphics[width= 1.0\columnwidth]{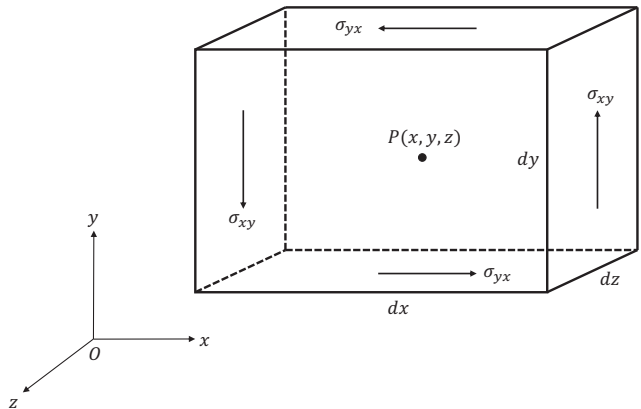}
\caption{ Torque produced by anti-symmetrical stresses on an elementary volume $dV=dxdydz$ relative to the center $P(x,y,z)$ of the volume. On the right positive $x$ surface, the force $\sigma_{xy}dydz$ has an arm $dx/2$. Summing up all four torques one obtains the torque $dM_{z}=2\sigma_{xy}dV=2c^{2}\left(\partial_{x}A_{y}-\partial_{y}A_{x}\right) dV$ in $z$ direction. Similarly for all other components of $d\textbf{M}$.}
\label{fig1}
\end{figure}

In order to understand this equation we consider an infinitesimal volume $dV$ of the 
aether and we calculate the torque using Eq. \eqref{stresstensor} acting on this volume (see Fig. \ref{fig1}). The torque is obviously $d\textbf{M}=2c^{2}\left(\nabla\times\textbf{A}\right)dV$ and therefore $-2c\textbf{B}=d\textbf{M}/dV$. We reach the conclusion that apart from a numerical factor, the magnetic field at a point is a torque acting per unit volume at this point. Having made these definitions we readily obtain Maxwell's equations in Lorentz-Heaviside system of units. Indeed $\nabla\cdot{\textbf{E}}=\nabla\cdot{\left(\rho\textbf{v}\right)}=-\dot{\rho}=\varrho$. Also $-\dot{\textbf{B}}=c\nabla\times\dot{\textbf{A}}=c\nabla\times\left(\rho \textbf{v}+\nabla\phi\right)=c\nabla\times \textbf{E}$. Obviously $\nabla\cdot \textbf{B}=-c\nabla\cdot{\left(\nabla\times \textbf{A}\right)}=0$. As for $\nabla\times \textbf{B}$ we derive,
\begin{equation}
\nabla\times \textbf{B}=-c\nabla\times\left(\nabla\times \textbf{A}\right)=\frac{\rho}{c}\dot{\textbf{v}}.
\end{equation}
We complete the full derivative and we obtain
\begin{equation}
\nabla\times \textbf{B}=\frac{1}{c}\frac{\partial}{\partial t}\left(\rho \textbf{v}\right)-\frac{1}{c}\dot{\rho}\textbf{v}=\frac{1}{c}\dot{\textbf{E}}+\frac{1}{c}\varrho \textbf{v}.\label{Maxwell4}
\end{equation}
Amazingly, in order to obtain the familiar Maxwell's equations we have to postulate that \textit{the charge moves with the velocity of the aether}, i.e.
\begin{equation}
\textbf{v}_{\text{charge}}(\textbf{r},t)=\textbf{v}(\textbf{r},t),\label{velocities}
\end{equation}
and with the additional definition of charge current $\textbf{J}=\varrho\textbf{v}_{\text{charge}}$ we finally have $c\nabla\times \textbf{B}=\textbf{J}+\dot{\textbf{E}}$. But why \eqref{velocities} should hold  true, unless\textit{ the charge is part of the aether, a kind of singularity in it.} This is quite an interesting consequence of the model. We also see (Eq. \eqref{model3}) that the aether is conserved and is not produced (or annihilated) by the charges.

\subsection{Some necessary qualifications}

The proposed theory is not \textit{yet} equivalent to Maxwell's equations. Indeed, $\textbf{E}$ and $\textbf{J}$ are always parallel in the theory proposed so far, since both are proportional to $\textbf{v}$. However we shall perform spatial averaging and coarse-graining. The new theory will become completely equivalent to Maxwell's theory of electrodynamics.

To this end we treat the spatial derivatives in our axioms (\ref{stresstensor}-\ref{model3}) as \textit{finite} differences over \textit{finite} volumes $\delta V$. These volumes are so small, that the finite differences will be \textit{approximately} the same as derivatives. In fact, we shall even write them and calculate them as derivatives. These small volumes $\delta V$ however, may contain many charges. This also shows that in the new theory the axioms (\ref{stresstensor}-\ref{model3}) will be considered as a kind of 'averaged' equations. They cannot be applied on the microscale (within $\delta V$) but on the macroscale only, i.e. over volumes much greater than $\delta V$. That is why we shall call the new theory 'macro-theory'.

Let us consider a volume $\delta V$ which has a center at the point $\textbf{r}$. We divide the volume $\delta V$ into $N$ equal cells. We assume that $N\gg 1$. Then $\textbf{E}(\textbf{r},t)$ at the point $\textbf{r}$ is defined in the macro-theory as a spatial average flux, i.e. 
\begin{equation}
\textbf{E}(\textbf{r},t)=\frac{1}{N}\sum_{k}\rho_{k}\textbf{v}_{k}\equiv\langle \rho\textbf{v}\rangle \label{electricfield-average},
\end{equation}
where the sum is spread through all cells in the small volume $\delta V$. All other quantities are defined as such spatial averages, i.e. $\textbf{J}=\langle\varrho\textbf{v}\rangle$, $\textbf{B}=-c\langle\nabla\times\textbf{A}\rangle$, etc. Now clearly 
\begin{equation}
\textbf{J}(\textbf{r},t)=\frac{1}{N}\sum_{a}\varrho_{a}\textbf{v}_{a}=-\frac{1}{N}\sum_{a}\dot{\rho}_{a}\textbf{v}_{a}.\label{current-average}
\end{equation}
The index $a$ shows that the sum is spread through \textit{only} those cells where there \textit{are} charges ($\dot{\rho}_{a}\neq 0$). By comparing Eqs. \eqref{electricfield-average} and \eqref{current-average} we see that $\textbf{E}$ is not in general parallel to $\textbf{J}$ (at the same point $\textbf{r}$) in the macro-theory.  

Next, we shall derive \textit{one} of Maxwell's equations within the framework of the macro-theory. All other Maxwell's equations can be derived in a similar way. To this end we take the time derivative of $\textbf{E}$:
\begin{equation}
\dot{\textbf{E}}=\frac{1}{N}\sum_{k}\dot{\rho}_{k}\textbf{v}_{k}+\frac{1}{N}\sum_{k}\rho_{k}\dot{\textbf{v}}_{k}=-\frac{1}{N}\sum_{a}\varrho_{a}\textbf{v}_{a}+\langle\rho\dot{\textbf{v}}\rangle.
\end{equation}\black
Now we can apply the axiom \eqref{model2} because we have already used coarse graining. However we omit the brackets $\langle\rangle$. Then,
\begin{equation}
\dot{\textbf{E}}=-\textbf{J}+c\nabla\times\textbf{B}.
\end{equation}\black
We have derived one of Maxwell's equations within the framework of the macro-theory. In a similar manner all other Maxwell's equations can be derived. We see that the coarse-graining is \textit{essential} in order to show that the charge motion cannot be derived from the axioms (\ref{stresstensor}-\ref{model3}). This is so because the axioms are true only on the macro-scale, not on the micro-scale. Therefore additional equations are necessary. These equations are Newton's equations for the charges (which we do not model here) in each cell, i.e. in the \textit{microscale}. \textit{Nevertheless, we still have} $\textbf{v}_{\text{charge}}=\textbf{v}$, i.e. the charges appear to be part of the aether.

In order to obtain the \textit{macroscopic} Newton's equations for the charges we take into account that Maxwell's equations could be derived from a Lagrangian density $\mathcal{L}=\mathcal{L}_{f}+\mathcal{L}_{fc}$, where $\mathcal{L}_{f}=\textbf{E}^{2}-\textbf{B}^{2}$ is the field Lagrangian density and $\mathcal{L}_{fc}=-\textbf{J}\cdot\textbf{A}$ is the coupling term. By simply adding the charge Lagrangian $L_{c}=-\sum_{\alpha}m_{\alpha}c^{2}\sqrt{1-\frac{\textbf{v}_{\alpha}^{2}}{c^{2}}}$ one easily derives Newton's equations for the charges $\dot{\textbf{p}}_{\alpha}=q_{\alpha}\textbf{E}(\textbf{r}_{\alpha},t)+q_{\alpha}\frac{\textbf{v}_{\alpha}}{c}\times\textbf{B}(\textbf{r}_{\alpha},t)$, where $\textbf{p}_{\alpha}$ is the relativistic momentum of particle $\alpha$.

The arbitrariness in the scalar potential (adding $\nabla\phi$ in Eq. \eqref{model1} does not alter the stress-tensor) is in fact the familiar gauge invariance of Maxwell's equations. Even the gauge $\phi=0$ can be used (the so called Weyl's gauge), which however is an incomplete gauge and people more often prefer Coulomb's gauge or Lorentz's gauge.

It is obvious that if there are no charges ($\varrho=0$, $\rho=\text{const.}$), we can integrate equation \eqref{model1} (choosing the gauge $\phi=0$) and obtain $\textbf{A}=\rho \textbf{u}$, where $\textbf{u}$ is the displacement of the aether. Then equation \eqref{model2} can be rewritten as $\dot{\textbf{v}}=-c^{2}\nabla\times(\nabla\times \textbf{u})$, which is MacGullagh's model of an aether that resists rotations rather than distortions since the torque $d\textbf{M}\sim\nabla\times \textbf{u}$ . The equation of motion becomes $\square\textbf{u}=0.$ Thus light becomes a kind of 'sound' wave in the aether.

\section{Statistical-mechanical interpretation of Lorentz transformations}
Major criticism against the model presented here may come from the familiar assumption that relativity theory dealt a final blow on any mechanical picture of electrodynamics. However this is not so.  In this section we \textit{derive} clock dilation and Fitzgerald-Lorentz contraction by starting with space and time according to Newton. We do not merely assume the distortions of the instruments as is usually done in LPI but we derive them. In addition, we show the importance to differentiate between what Lorentz called 'local time' (which is merely a good notation) and the false reading of the clock ($t^{\prime}$). Unless one stresses on this difference, a great deal of confusion ensues (it is for this reason that we repeat in detail the familiar derivation of Lorentz covariance of the wave equation). In addition we show that Lorentz covariance should always be made \textit{not only} of the wave equation (Maxwell's equations) but \textit{also} for equations describing the motion of the matter (the charges). In this section we examine the Lorentz covariance of the \textit{system} of Maxwell's equations \textit{and} Boltzmann's equation. \black Lorentz covariance of Boltzmann's equation was proved by Clemmow and Wilson \cite{Clemmow1956} within Minkowskian interpretation but the important consequences for LPI have not been considered. One of them is that we may show the specific mechanism of how the clocks are delayed and the rods are contracted. \black To show that, we derive Lorentz transformations by investigating the one-particle distribution $f(\textbf{r},\textbf{p},t)$ of a gas of molecules. We take into account that the molecules of the gas emit electromagnetic field and this field acts \textit{back} on the molecules themselves and thus distorts $f(\textbf{r},\textbf{p},t)$. It turns out that in order for a gas in motion to remain in thermal equilibrium the gas becomes contracted with the standard FitzGerald-Lorentz contraction. We show that clock dilation results for similar reasons. This derivation also shows that the increase of the life-time of elementary particles (say muon) may be explained by the presence of some inner structure of these particles. \black

We start by assuming Newtonian space and time and a privileged aether frame of reference. \black Second, we consider a gas in the aether with both positive and negative charges. Therefore we use two probability distributions, which we call $f_{j}$, $j=1,2$.  Boltzmann's equations for both are
\begin{equation}
\frac{\partial f_{j}}{\partial t}+\textbf{v}\cdot\frac{\partial f_{j}}{\partial \textbf{r}}=e_{j}\left(\textbf{E}+\frac{\textbf{v}}{c}\times\textbf{B}\right)\cdot\frac{\partial f_{j}}{\partial \textbf{p}},
\end{equation}
where $e_{j}=\pm e$ are the charges of the molecules. By using the standard Lorentz gauge $\nabla\cdot\textbf{A}-\dot{\phi}/c^{2}=0$, the field equations are 
\begin{equation}
\square \phi=-\varrho,
\end{equation}
and $\square\textbf{A}=\textbf{J}/c^{2}$. If there are $\mathcal{N}$ positive and negative charges, the charge and current density become
\begin{align}
&\varrho=e\mathcal{N}\int f_{1}(\textbf{r},\textbf{p},t)d^{3}\textbf{p}-e\mathcal{N}\int f_{2}(\textbf{r},\textbf{p},t)d^{3}\textbf{p},\\
& \textbf{J}=e\mathcal{N}\int \textbf{v}f_{1}(\textbf{r},\textbf{p},t)d^{3}\textbf{p}-e\mathcal{N}\int \textbf{v}f_{2}(\textbf{r},\textbf{p},t)d^{3}\textbf{p}.
\end{align}
We finally have the \textit{system} Maxwell+Boltzmann equations, which describes the gas. \black

\subsection{A gas in thermal equilibrium moving relative to the aether. New notation for the Maxwell's wave equations}

Let us begin by considering a gas in \textit{thermal equilibrium} at rest relative to the aether. Obviously then the probability distributions are time-\textit{in}dependent. Let us use superscript $(0)$ to denote the gas \textit{at rest} with respect to the aether. Thus, the equations describing this gas in absolute rest and at the same time in thermal equilibrium are,
\begin{equation}
\nabla^{2}\phi^{(0)}=e\mathcal{N}\int f_{1}^{(0)}(\textbf{r},\textbf{p})d^{3}\textbf{p}-e\mathcal{N}\int f_{2}^{(0)}(\textbf{r},\textbf{p})d^{3}\textbf{p}.
\end{equation}
Similarly for $\textbf{A}^{(0)}$. Boltzmann's equations are
\begin{equation}
\textbf{v}\cdot\frac{\partial f_{j}^{(0)}}{\partial\textbf{r}}= e_{j}\left(\textbf{E}^{(0)}+\frac{\textbf{v}}{c}\times\textbf{B}^{(0)}\right)\cdot\frac{\partial f_{j}^{(0)}}{\partial\textbf{p}}.
\end{equation}
We shall assume that the solution of any gas at rest and in thermal equilibrium can be obtained, more precisely we can obtain $\phi^{(0)}$, $\textbf{A}^{(0)}$ and $f_{j}^{(0)}$.

Now, let us examine \textit{the same} gas in \textit{thermal equilibrium}, which however moves with velocity $V$ in $x$ direction with respect to the aether. We do not change the reference frame. The frame is still the aether frame. The equation describing the potential for this moving gas is
\begin{equation}
\square\phi=-\varrho(x-Vt,y,z).\label{PhiEquil}
\end{equation}
Similarly for $\textbf{A}$. The only time dependence is due to the general motion of the gas with respect to the aether. Let us substitute
\begin{align}
&x_{1}=x-Vt,\notag\\
&y_{1}=y,\notag\\
&z_{1}=z,\notag\\
&t_{1}=t.\label{Coordinates1}
\end{align}
Please note that this is simply a \textit{notation}. \textit{No physical meaning whatsoever} is given to these substitutions. The wave equation for the potential $\phi$ is changed to,
\begin{equation}
\left(1-\frac{V^{2}}{c^{2}}\right)\frac{\partial^{2}\phi}{\partial x_{1}^{2}}+\frac{\partial^{2}\phi}{\partial y_{1}^{2}}+\frac{\partial^{2}\phi}{\partial z_{1}^{2}}+\frac{2V}{c^{2}}\frac{\partial^{2}\phi}{\partial x_{1}\partial t_{1}}=-\varrho(x_{1},y_{1},z_{1}).\label{Eq1}
\end{equation}
Similarly for $\textbf{A}$. On the right hand side, the charge is time-independent in the new notation. However the wave equation on the left is distorted. It is obvious though that the left hand side will be simplified if we substitute
\begin{align}
&x_{2}=\frac{x_{1}}{\sqrt{1-\frac{V^{2}}{c^{2}}}}=\frac{x-Vt}{\sqrt{1-\frac{V^{2}}{c^{2}}}},\notag\\
&y_{2}=y_{1}=y,\notag\\
&z_{2}=z_{1}=z,\notag\\
&t_{2}=t_{1}=t. \label{Coordinates2}
\end{align}
Then Eq. \eqref{Eq1} becomes
\begin{equation}
\square_{2}\phi+\frac{2V}{c^{2}}\frac{\partial^{2}\phi}{\partial x_{2}\partial t_{2}}=-\varrho\left(x_{2}/\gamma,y_{2},z_{2}\right),\label{Eq2}
\end{equation}
$\square_{2}$ is the D'Alambert operator with respect to the notation \eqref{Coordinates2} and $\gamma = \left(1-V^{2}/c^{2}\right)^{-1/2}$ Eq. \eqref{Eq2} however is not a wave equation, even though on the right hand side we have a time-\textit{in}dependent source. It is not difficult however to guess a method to correct that. We substitute,
\begin{align}
& x_{3}=x_{2}=\gamma\left(x-Vt\right),\notag\\
& y_{3}=y_{2}=y,\notag\\
& z_{3}=z_{2}=z,\notag\\
& t_{3}=t_{2}\gamma-x_{2}\frac{V}{c^{2}}=\gamma\left(t-xV/c^{2}\right).\label{Coordinates3}
\end{align}
In this manner we finally obtain,
\begin{equation}
\square_{3}\phi=-\varrho\left(x_{3}/\gamma,y_{3},z_{3}\right).
\end{equation}
This is truly an inhomogeneous wave equation with a time-\textit{in}dependent source. Again we stress that Eqs. \eqref{Coordinates3} are simply a notation. Nothing else. The same considerations are done for the vector potential $\square_{3}\textbf{A}=\frac{1}{c^{2}}\textbf{J}\left(x_{3}\gamma,y_{3},z_{3}\right)$.
One might think that the job to convert the equations of a moving gas in an effective gas at rest is done. However we are not ready because for these new equations the sources on the right hand side do not obey the continuity equation. Indeed, if we rewrite the continuity equation
\begin{equation}
\frac{\partial\varrho}{\partial t}+\nabla\cdot\textbf{J}=0,
\end{equation}
with the new notation \eqref{Coordinates3}, we have,
\begin{equation}
\frac{\partial}{\partial t_{3}}\gamma\left(\varrho-VJ_{x}/c^{2}\right)+\frac{\partial}{\partial x_{3}}\gamma\left(J_{x}-V\varrho\right)+\frac{\partial}{\partial y_{3}}J_{y}+\frac{\partial}{\partial z_{3}}J_{z}=0.
\end{equation}
In order for the new sources to obey the continuity equation we have to perform another substitution
\begin{align}
&\varrho_{3}=\gamma\left(\varrho-VJ_{x}/c^{2}\right),\\
&J_{3x}=\gamma\left(J_{x}-V\varrho\right),
\end{align}
and $J_{3y}=J_{y}$ and $J_{3z}=J_{z}$. We have to perform similar linear combinations of $\phi$ and $\textbf{A}$ such that the sources of the wave equations are $\varrho_{3}$ and $\textbf{J}_{3}$.
\begin{align}
&\phi_{3}=\gamma\left(\phi+VA_{x}\right),\notag\\
&A_{3x}=\left(A_{x}+V\phi/c^{2}\right),
\end{align} 
and $A_{3y}=A_{y}$ as well as $A_{3z}=A_{z}$. Then we finally have
\begin{equation}
\square_{3}\phi_{3}=-\varrho_{3}\left(x_{3}/\gamma,y_{3},z_{3}\right).
\end{equation}
Similarly for $\textbf{A}_{3}$. The wave equations are finally reduced to equations with stationary sources. The latter obey the continuity equation
\begin{equation}
\frac{\partial\varrho_{3}}{\partial t_{3}}+\nabla_{3}\cdot\textbf{J}_{3}=0.
\end{equation}
Since the sources in these wave equations in the new notation are \textit{effectively} time-independent, we may discard the time derivative
\begin{equation}
\nabla^{2}_{3}\phi_{3}=-\varrho_{3}\left(x_{3}/\gamma,y_{3},z_{3}\right).\label{Field3}
\end{equation}
Similarly for $\textbf{A}_{3}$. Next, we consider Boltzmann's equations.

\subsection{The new notation applied to Boltzmann's equations for the moving gas in thermal equilibrium}
Boltzmann's equations for the moving gas in thermal equilibrium for both type of charges are,
\begin{equation}
\frac{\partial f_{j}}{\partial t}+\textbf{v}\cdot\frac{\partial f_{j}}{\partial \textbf{r}}=e_{j}(\textbf{E}+\frac{\textbf{v}}{c}\times\textbf{B})\cdot\frac{\partial f_{j}}{\partial \textbf{p}}.\label{BoltzmannEquil}
\end{equation}
However we have $f_{j}=f_{j}(x-Vt,y,z,p_{x},p_{y},p_{z})$. The only time dependence of $f_{j}$ which describes a thermal equilibrium of a moving gas is due to general locomotion of the gas. We use directly substitution  \eqref{Coordinates3}. In order to rewrite Boltzmann's equations we have to correct the electromagnetic field using $\textbf{E}_{3}$ and $\textbf{B}_{3}$ derived from the potentials $\phi_{3}$ and $\textbf{A}_{3}$. In addition, we have to distort the axes in the phase space. The coordinates $x,y,z,t$ are distorted according to Eqs. \eqref{Coordinates3}. But we need to do the same for the moment. To this end we have
\begin{equation}
v_{3x}=\frac{dx_{3}}{dt_{3}}=\frac{v_{x}-V}{1-\frac{v_{x}V}{c^{2}}}.
\end{equation}
Similarly, one could derive $v_{3y}$ and $v_{3z}$ and from here the momentum $\textbf{p}_{3}$ which allows us to distort all axes in phase space using Eqs. \eqref{Coordinates3}. After a somewhat laborious calculation (the reader may also follow Clemow and Wilson \cite{Clemmow1956} for a four-dimensional notation) we obtain,
\begin{equation}
\textbf{v}_{3}\cdot\frac{\partial f_{j}}{\partial \textbf{r}_{3}} =e_{j}\left(\textbf{E}_{3}+\frac{\textbf{v}_{3}}{c}\times \textbf{B}_{3}\right)\cdot\frac{\partial f_{j}}{\partial \textbf{p}_{3}}.
\end{equation}
We see that Boltzmann's equation is covariant with respect to the new notation Eq. \eqref{Coordinates3}. This equation, being combined with the effective stationary field equation \eqref{Field3}, and a similar Poisson's equation for $\textbf{A}_{3}$ helps us to reach the conclusion that the whole of the moving gas is \textit{effectively} reduced to a gas which is in absolute rest (i.e. rest relative to the aether). Of course, the gas is in fact moving with velocity $V$ along the $x$ axis but with the new notation, the gas behaves \textit{mathematically} as if it is in absolute rest. Reverting back to the original notation we have
\begin{align}
&f_{j}(\textbf{r},\textbf{p},t)=\notag\\
&f_{j}^{(0)}\left(\gamma\left(x-Vt\right),y,z,\gamma\left(p_{x}-V\sqrt{{\textbf{p}}^{2}c^{2}+m^{2}c^{4}}/c^{2}\right),p_{y},p_{z}\right)\label{SolutionEquilibrium}
\end{align}
The solution for the distributions $f_{j}$ is rewritten in terms of some effective system which is now effectively in absolute rest. However, the above rest solution $f_{j}^{(0)}$ is \textit{not any} effective solution but is the solution of the \textit{particular gas which is in motion}, if it were \textit{not} in motion but if it were in absolute rest. This is immediately seen by taking the limit $V\rightarrow0$ on the right hand side of Eq. \eqref{SolutionEquilibrium}. 

In that way we see that the initial distributions $f_{j}^{(0)}$ are translated along the $x$-axis with amount $Vt$ and then contracted with a factor $\gamma^{-1}$. \textit{Therefore the whole gas is contracted and this leads to the familiar FitzGerald-Lorentz contraction even though we used Newtonian space and time}. It is easy to see that the cause of this contraction is the electromagnetic field created by the molecules of the gas and which acts back upon the gas. Indeed, we have examined a complicated self-interacting system of molecules within Newtonian time and space and we have made no relativistic assumptions. Not only that but we know that the field equations for $\phi$ and $\textbf{A}$ are Lorentz covariant. If they were Galilean covariant, then it is easy to see that Boltzmann's equations would have been Galilean covariant as well, and the gas would \textit{not} have contracted.

\subsection{Dilation of all processes in a body \textbf{not} in thermal equilibrium}
We can also examine a time-dependent situation for a gas, which is \textit{not} in thermal equilibrium. In exactly the same way as for a gas in thermal equilibrium we can show that the distribution in this case is:
\begin{align}
f_{j}(\textbf{r},\textbf{p},t)=f_{j}^{(0)}\left(\gamma\left( x-Vt\right),y,z,\gamma\left(t-xV/c^{2}\right),...\right).\label{FinalDistribution}
\end{align}
We observe that the gas is again FitzGerald-Lorentz contracted. However, we also see that all processes are delayed with the factor $\gamma$, and with the amount $\gamma xV/c^{2}$ depending on $x$. \textit{Thus we have derived the familiar clock delay within Newtonian space and time}. \black This effect is also due to back-action of the electromagnetic field.

\subsection{Another reference frame moving with respect to the aether using physical clocks and physical rods as measuring instruments}
So far, we have shown how a gas of particles gets slowed down and is contracted. This 'proof of principle' calculation shows the mechanism of distortion of any material body, including the measuring instruments (clocks and rods) in a reference frame. \black Thus are finally ready to examine what is going to happen if a new reference system $K^{\prime}$ is used, which moves relative to the aether $K$ with velocity $V$ in $x$ direction. We assume that the centers of the two coordinate systems $O$ and $O^{\prime}$ coincide at $t=0$. In this new reference system however, we must take into account that the clocks and rods which we use to measure time and distance are physical devices, i.e. they are made of molecules and therefore they are themselves distorted. All physical rods are contracted and all clocks are delayed, similarly to the gas in the previous subsections. 

Let an event $M$ occurs in a point $(x,y,z)$ in the aether at absolute moment of time $t$. What are the coordinates in the new reference system $K^{\prime}$? Obviously, if there was no FitzGerald-Lorentz contraction, then $x^{\prime}=x-Vt$. However, since the moving measuring rods are contracted they measure greater distance $x^{\prime}=\gamma\left(x-Vt\right)$. Please take into account the difference between $(x^{\prime},y^{\prime},z^{\prime})$ and $(x_{3},y_{3},z_{3})$ in Eqs. \eqref{Coordinates3}. The latter are mere notation which was useful to obtain Eq. \eqref{FinalDistribution}. However, the coordinates $(x^{\prime},y^{\prime},z^{\prime})$ are \textit{not} notation but rather the \textit{false} reading of the \textit{distorted} measuring rods in a \textit{moving} reference system.
In exactly the same way we reach the conclusion that the false reading of the distorted clocks is $t^{\prime}=\gamma\left(t-Vx/c^{2}\right)$. Again, note the difference between $t_{3}$ in Eqs. \eqref{Coordinates3} and $t^{\prime}$. Here $t_{3}$ is a mere notation which was helpful to establish Eq. \eqref{FinalDistribution}, and which Lorentz called 'local' time.  However $t^{\prime}$ is the measure of the distorted clocks in a moving reference frame $K^{\prime}$. It is Einstein's greatest achievement that he went beyond the local time $t_{3}$ and introduced $t^{\prime}$.

From these Lorentz transformations for $x^{\prime},y^{\prime},z^{\prime}$ and $t^{\prime}$ we can derive all familiar results. That all inertial reference frames \textit{appear} to be indistinguishable and equivalent, that the speed of light $c$ \textit{appears} to be invariant in all inertial reference frames. However in LPI, this appearance is just that - an \textit{appearance} which is however false and is due to the distortion of the measuring instruments of the moving reference system.

We have made a \textit{dynamical} derivation of Lorentz transformation, not kinematical. The debate between these two points of view on the nature of Lorentz transformation is onging and quite interesting.\cite{Maudlin2012, Brown2005} \black

In addition, we can explain why the muon has a greater life time when it moves with greater speed. The explanation from the point of view of the aether is that the muon has some internal structure. There are internal forces inside the muon (not necessarily electromagnetic), which propagate with finite speed and distort the muon and increase its life time.

Lastly, even if quantum mechanical considerations were to be applied and even if some collision terms included in the Boltzmann's equations, so long as the Lorentz covariance is applicable then the gas would be distorted. Therefore the results are quite general.

\section{What's next}
We have shown the possibility for a mechanical model of the electromagnetic field, even though it is Lorentz covariant and without violation of the conservation of the aether. The next natural step is to attempt a mechanical model of Einstein's gravity equations within LPI. This task does not seem impossible at all. Indeed, let us confine our considerations with the linearized Einstein's gravity theory \textit{without} matter. In the absence of matter the aether's density $\rho = \text{const.}$ and $\textbf{A}=\rho\textbf{u}$. If we postulate $h_{00}=\rho c^{2}$, $h_{0i}=\rho c v_{i}$ and $h_{ij}=-\sigma_{ij}$, Newton's equations $\rho\dot{v}_{i}=\partial_{k}\sigma_{ki}$ and the continuity equation $\dot{\rho}+\nabla\cdot\left(\rho \textbf{v}\right)=0$ for the aether can be written in a four dimensional notation as $\partial_{\mu}h^{\mu\nu}=0$. This equation corresponds to the gauge condition in gravity theory. Obviously we also have that $\square h_{\mu\nu}=0$. In this way we have derived, using the model's equations for the aether, the linearized Einstein's graivity equations. Wyss \cite{Wyss1965} has constructed iterative procedure to derive Einstein's gravity equations from the linear theory. Then Deser \cite{Deser1970} improved the tecnhique. Independenly Thirring \cite{Thirring1961} has shown that a particle interacting with this tensor field, will move as if in a metric $g_{\mu\nu}=\eta_{\mu\nu}+\varepsilon h_{\mu\nu}$ to first order in $h_{\mu\nu}$, while $\eta_{\mu\nu}$ is the true metric (according to LPI $\eta$ is only an instrument), which is flat. In LPI we may say that due to the distortions of the instruments by the gravity field the true metric $\eta_{\mu\nu}$ will be concealed and instead instruments will observe $g_{\mu\nu}$. A full aether theory would require some strain-stress relations and should incorporate the distortion of the instruments at each point in space and time similarly to Arminjon's scalar aether theory \cite{Arminjon2004}, which is able to reproduce Schwarzschild's metric.

\section{Acknowledgment}
I would like to thank to Ivaylo Papazov for helpful discussions.

This research did not receive any specific grant from funding agencies in the public, commercial, or not-for-profit sectors. 

\section{Data Availability Statement}
The author confirm that the data supporting the findings of this study are available within the article and/or its supplementary materials.

\appendix

\section{A and B-theory of time}
In order to understand the differences between the three interpretations of relativity theory, one needs first to take into account that there are two models of time \cite{Craig2001}, called tensed theory of time (also A-theory of time) and tense\textit{less} theory of time (also B-theory of time).

According to A-theory of time only the present is real (i.e. only the present exists), the future does \textit{not} exist (it \textit{will} exist) and the past does \textit{not} exist (it \textit{no longer} exists). This is the common sense notion of time. Let us imagine a staircase and let each stair represents a moment of time. According to A-theory of time one particular stair (present) exists, the stairs below this stair (past) no longer exist and the stairs above this stair (future) do not exist yet. When the next moment of time comes (and \textit{it} becomes present), it comes into being and the previous stair (which becomes past) ceases to exist. Such is the classical notion of the \textit{flow} of time.

According to B-theory of time, the whole staircase exists and is real, i.e. not only the present exists but also the past and the future. The flow of time is a subjective illusion in B-theory. Such a model of time allows the hypothetical possibility of going back in time (getting down to lower stairs), while the A-theory of time does not allow this possibility (since the past does not exist). The Minkowskian interpretation as we shall see below rests on the assumption of B-theory of time.

\section{Lorentz-Poincare interpretation}
Lorentz-Poincare interpretation starts with the notions of space and time according to Newton. This means that there exists an \textit{absolute} time and \textit{absolute} space. Absolute time flows uniformly of its own nature and without reference to anything external. It is different than the \textit{physical} time, which is the \textit{measure} of absolute time by \textit{physical} clocks and material bodies. The physical clocks are delayed when in motion but \textit{not} the absolute time. The same with space. According to Newton there are two kinds of spaces - absolute and physical. Absolute space is homogeneous and immovable, it exists without reference to anything external. However, \textit{physical} space measured by physical processes (light signals or physical rods) is merely a measure of the absolute space. This distinction shows immediately that there exists a special reference frame which should give physical time and physical space in coincidence with absolute time and absolute space.

According to LPI Lorentz transformations merely describe how \textit{physical} rods and clocks are contracted and delayed when in motion. They connect reference frames made by physical rods and clocks amenable to alteration when in motion.

LPI has been further developed (neo-Lorentzian interpretation) to as few as possible assumptions. In fact it has become as simple as the relativistic and Minkowskian interpretations. 

This interpretation gives us physical \textit{causes} for the clock dilation and rod contraction, namely physical \textit{forces}. Electromagnetic force literally acts on the arrow of the clock and slows it down. The interpretation assumes A-theory of time and standard notion of the flow of time.

\section{Relativistic interpretation}
The second interpretation (the relativistic interpretation) is the original Einstein's interpretation of his 1905 paper \cite{Einstein1905}. In this interpretation space-time is merely an instrument, a helpful tool, and is not interpreted realistically (as in Minkowskian interpretation). It assumes A-theory of time. Einstein dropped this interpretation later in favor of the Minkowskian interpretation.

It is customary to present this interpretation in terms of the postulate of the relativity of all inertial reference frames \textit{and} the postulate of the constancy of the velocity of light $c$. However, in order to define properly the meaning of the words 'reference frame', one needs more preparatory work and we shall see that instead of two axioms, we need in fact eleven, eight of which are mere conventions (at least in this interpretation) and three are empirical. We shall follow Reichenbach \cite{Reichenbach1957}.

\subsubsection{The relativity of the simultaneity of distant events}

\begin{figure}[tb]
\includegraphics[width= 0.5\columnwidth]{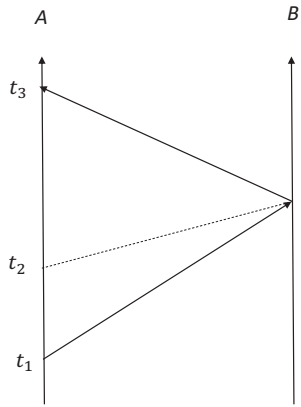}
\caption{Clock synchronization according the relativistic interpretation. Light signal is sent from clock $A$ (the vertical axis is the time axis) to clock $B$ and reflected back to $A$. The moment $t_{2}$ (as measured by $A$) when the signal reached $B$ is chosen by convention, i.e. $t_{2}=t_{1}+\epsilon\left(t_{3}-t_{1}\right)$ for any $\epsilon$ ($0<\epsilon <1$). Einstein has chosen the convention $\epsilon=\frac{1}{2}$.}
\label{fig2}
\end{figure}

We can easily establish whether two events at \textit{different} locations are simultaneous if there were infinitely fast signals. We simply send such a signal from point $A$ at moment of time $t_{1}$ to point $B$ and reflect it back to $A$. Clearly this signal, being infinitely fast, returns back to $A$ again at the moment $t_{1}$. However we do not have such signals, and thus absolute simultaneity of such a type cannot be established. Therefore we can use the fastest possible signal - light signal and send it from point $A$ at a moment of time $t_{1}$ as shown in Fig. \ref{fig2}. Then the light signal reaches point $B$ and is reflected back to $A$. It returns to $A$ at moment $t_{3}$. What moment of time $t_{2}$ measured by a clock in $A$ is simultaneous with the event when the light signal reached $B$? Obviously $t_{1}<t_{2}<t_{3}$. But since there are no infinitely fast signals, nor there are signals faster than the light signal, then it is impossible even in principle (according to the relativists) to establish $t_{2}$. Thus, the relativist claims that the moment $t_{2}$ is chosen by convention! In other words $t_{2}=t_{1}+\epsilon\left(t_{3}-t_{1}\right)$ and we can choose by convention any $\epsilon$ such that $0<\epsilon<1$. The choice $\epsilon=\frac{1}{2}$ is one such possibility. If we choose $\epsilon=\frac{1}{2}$ it appears that we have assumed that the light signal travels in both directions with the same speed. However this is \textit{not} true. We have in fact \textit{defined} it to travel in both directions with the same speed. The choice of any $\epsilon$ is a convention and it defines simultaneity of distant events. Therefore the constancy of the speed of light in both directions (being the fastest signal) is a convention, not an empirical fact. 

\subsubsection{Definition of reference frames. Lorentz transformations}

Let us imagine a continuum of points in the whole of space, each endowed with an observer. Let us consider a particular point $A$. The observer at $A$ defines his unit of time by some periodic process and let this unit of time be the second. 

\textit{Axiom 1: (convention):} Time flows uniformly at all points in space.

Next, the observer sends a light signal to some point $B$ and reflects it back to $A$. Let us denote with $\overline{ABA}$ the time interval for the whole trip of the light signal $A-B-A$ as measured by the clock at $A$.

\textit{Axiom 2 (convention)}: If the point $B$ has the property that the time interval $\overline{ABA}$ is always the same as measured by a clock at $A$, no matter when the light signal is sent from $A$, we \textit{define} such a point of being at \textit{rest} relative to $A$. 

Please note that this is a mere convention and in fact a definition of rest. Now, the observer at $A$ finds other points $C$, $D$, etc. being at rest relative to $A$. We call such a \textit{system} of points at rest relative to $A$. However, just because $\overline{ABA}=\text{const.}$, $\overline{ACA}=\text{const.}$, etc. it does \textit{not} follow that $\overline{BAB}=\text{const.}$ or $\overline{CAC}=\text{const}$. In other words, the points $B$, $C$, etc. are at rest relative to $A$ but it does \textit{not} follow that $A$ is at rest relative to $B$ or to $C$ or to any other point. That such systems of points exist with the special property that all points are at rest relative to each other is an empirical fact (we do not consider general relativity here). 

\textit{Axiom 3 (empirical fact):} There exist special systems of points $A$, $B$, $C$,..., such that all points are at rest relative to each other.

Note, there is not just a single system of points but infinite such systems.

\textit{Axiom 4: (convention):} We select such a system of points which are at rest relative to each other.

Next, the observer at $A$ sends his time unit (second) to the other observers at $B$, $C$, etc. He may do so by merely sending light signals every second. Please note that the unit of time is thus transferred to the other observers, but clocks are \textit{not yet} synchronized, i.e. the notion of simultaneity of distant events is not established yet.

Let us choose three points, $A$, $B$ and $C$ of our selected system of points. Therefore these points are at rest relative to each other. And let us send \textit{two} signals \textit{simultaneously} from $A$. One of the signal travels the trip $A-B-C-A$ and the other $A-C-B-A$. Now, \textit{generally} the two signals will \textit{not} return to the point $A$ simultaneously (measured by the clock at $A$) even though the points $A$, $B$ and $C$ may be at rest relative to each other. That there exist such systems of points that the round trip journey takes the same amount of time is an empirical fact (again, we exclude general relativity here).

\textit{Axiom 5: (empirical fact)} There exist special systems of points, at rest relative to each other such that the round-trip journeys $\overline{ABCA}=\overline{ACBA}$ are always the same.

We are finally ready to define the simultaneity of distant events by light signal synchronization.

\textit{Axiom 6: (convention)} Distant clocks are synchronized using light signals. In other words if we choose two arbitrary points $A$ and $B$ of our selected system of points which are at rest relative to each other, we send a light signal at a moment of time $t_{1}$ measured by the clock at $A$. It travels the distance $A-B-A$ and returns at $A$ at a moment of time $t_{3}$ by the clock at $A$. The moment $t_{2}$ at $A$ simultaneous with the moment at $B$ when the signal reached $B$ is \textit{defined} to be $t_{2}=t_{1}+\epsilon\left(t_{1}-t_{3}\right)$ for $\epsilon=\frac{1}{2}$. In this manner clock $B$ is synchronized \textit{by} the clock in $A$. The clocks in all other points can be synchronized by the clock at $A$ in the same way.

The above definition may seem to have chosen a special point $A$. But it can be easily proved that the above synchronization procedure is symmetric. This means that the point $A$ is not special in any way and in fact if we were to choose any other point to synchronize all clocks, both synchronizations will agree, provided we choose the same $\epsilon$ (in our case by convention $\epsilon=\frac{1}{2}$) . In addition this synchronization is transitive, i.e. if two clocks at different points $B$ and $C$ are synchronized by $A$ they are synchronized by each other.

Thus far we have dealt with the concept of time in our selected system of points. Now we continue with space. The first notion is the topological notion of \textit{between}.

\textit{Axiom 7:(convention):} If we choose three points $A$, $B$ and $C$ in our selected system of points we define point $B$ to be \textit{between}  $A$ and $C$ if $\overline{ABC}=\overline{AC}$.

\textit{Axiom 8 (empirical fact):} If points $B_{1}$ and $B_{2}$ are between $A$ and $C$, then either $B_{2}$ is between $A$ and $B_{1}$ \textit{or} $B_{2}$ is between $B_{1}$ and $C$.

The above two axioms help us to define the notion of straight line.

\textit{Axiom 9 (convention):} The straight line through $A$ and $B$ is the set of all points which among themselves satisfy the relation \textit{between} and which include the points $A$ and $B$.

With this preparation in hand, we can define the equality of distances in our selected system of points.

\textit{Axiom 10 (convention):} If the time interval $\overline{ABA}=\overline{ACA}$ for three different points $A$, $B$ and $C$ in our selected system of points, then we define $|AB|=|AC|$.

This concludes the geometry of space. The above axioms are quite sufficient to prove that space becomes Euclidean.

\textit{Axiom 11 (convention):} Let us choose two inertial systems $K$ and $K^{\prime}$ as defined by the above axioms in different states of motion. Let $l$ be a rest-length in a system $K$ and $l^{\prime}$ be a rest-length in $K^{\prime}$. If $l$ is measured by observes at rest in $K^{\prime}$, they will not in general measure the same length $l$ as observers at rest in $K$. There will be some expansion or contraction factor. The same principle is true if $l^{\prime}$ is measured by observers at rest in $K$. We require by convention the \textit{identity} of these expansion (or contraction) factors obtained by the observes at rest in $K$ and $K^{\prime}$.

With these eleven axioms at our disposal we finally have a correct meaning of the notion of reference frame. Obviously the above axioms define the light signal to have the same velocity in each reference frame. Not only that but the geometry is Euclidean (we are still in special relativity) and the distance traveled by a light signal from point $(x,y,z)$ to point $(x+dx,y+dy,z+dz)$ is $c^{2}dt^{2}=dx^{2}+dy^{2}+dz^{2}$, where the right hand-side is the distance between two infinitesimally close points and $dt$ is the time required for the light signal to traverse that distance. In another reference frame we have the same speed, thus $c^{2}dt^{\prime 2}=dx^{\prime 2}+dy^{\prime 2}+dz^{\prime 2}$. \textit{Given our axioms}, the only transformations between $x,y,z,t$ and $x^{\prime},y^{\prime},z^{\prime},t^{\prime}$ that obey the above two equations simultaneously are the familiar Lorentz transformations. All familiar results follow from here.

Imagine a rod placed in $x$ direction in a reference frame $K$ and let it move with a velocity $V$ along $x$ direction relative to $K$. How is the length of the rod measured? One simply places two observers at \textit{some moment of time} $t$ (in $K$) placed at both ends of the rod and measures the distance between the observers. However, if one performs the same experiment in a reference frame $K^{\prime}$ which moves with the rod (i.e., the rod is at rest relative to $K^{\prime}$) the very notion of the same moment of time $t^{\prime}$ in $K^{\prime}$ is quite different than that in $K$ and thus different length is measured. Therefore the difference of the length of an object in different reference frames is connected with the relativity of simultaneity in different reference frames (according to the relativists).

The interpretation uses A-theory of time. This concludes the relativistic interpretation. 

\section{Minkowskian interpretation}
Minkowskian interpretation unites time and space into a four-dimensional manifold, called space-time. The space-time is not merely a helpful instrument but is interpreted realistically. The physical objects are four-dimensional. This interpretation assumes B-theory of time. The four dimensional distance between two points $(x,y,z,t)$ and $(x+dx,y+dy,z+dz,t+dt)$ in space-time is $ds^{2}=c^{2}dt^{2}-dx^{2}-dy^{2}-dz^{2}$.
The geometry in space-time is thus defined, as being pseudo-Euclidean geometry. Going from one inertial reference frame to another is again given by Lorentz transformations, but they are here interpreted as a change of coordinates in the space-time manifold.

\section{Assessment of the three interpretations}
We shall examine carefully the various interpretations of relativity theory.

\begin{figure}[tb]
\includegraphics[width= 1.0\columnwidth]{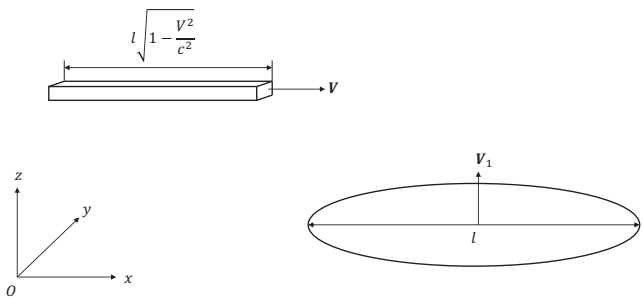}
\caption{The reality of FitzGerald-Lorentz contraction. As the rod, with rest-length $l$ moves with velocity $\textbf{V}$ in $x$ direction it is shortened to $l\sqrt{1-V^{2}/c^{2}}$. On the other hand, the diameter $l$ of the metal ring which moves with velocity $\textbf{V}_{1}$ in $z$ direction is \textit{not}. Therefore the rod \textit{can pass through} the ring!}
\label{fig3}
\end{figure}

We start with the relativistic interpretation. Let us imagine two objects\cite{Craig2001} - a rod and a metal ring in a reference frame $K$ in the configuration shown in Fig. \ref{fig3}. If there were no Lorentz contraction, the rod would not have been able to pass through the ring because its length is equal to that of the diameter of the ring. However, due to FitzGerald-Lorentz contraction the rod is shortened to $l\sqrt{1-V^{2}/c^{2}}$ while the diameter of the metal ring is not changed, since the velocity $\textbf{V}_{1}$ is perpendicular to the plane of the ring. Therefore, the rod will be able to pass through the metal ring! Of course, if one examines what happens from the reference frame of the rod, it is trivial to show that the ring will be inclined due to Lorentz contraction and the rod will still pass the ring. However in $K$ we see that the rod passes through the ring and so the FitzGerald-Lorentz contraction is a real physical phenomenon, not simply a result of the relativity of simultaneity as claimed by the relativists.

Let us examine another famous example - Bell's spaceship paradox\cite{Craig2001}. Two spaceships moving with the same velocity in an inertial reference frame $K$. Therefore the distance $L$ between them remains constant as they move. If these spaceships accelerate simultaneously (in $K$) with the same acceleration, then the distance between the spaceships obviously will remain the same $L$ even after they accelerate. Now, let us consider this scenario again but this time let us imagine a delicate string or thread that hangs between the spaceships, i.e. the string has a length $L$. Now, if the ships accelerate again with the same acceleration in $K$ the string will be subjected to FitzGerald-Lorentz contraction, i.e. its length will tend to be less than $L$, while the distance between the ships remains $L$ and the string will break! That it will break can be seen from the momentary inertial frame of the spaceships $K^{\prime}$, where due to the relativity of simultaneity the ships will not begin their acceleration simultaneously even though they accelerate simultaneously in $K$. Therefore FitzGerald-Lorentz contraction can break delicate strings.

Both of these scenarios can be multiplied \cite{Craig2001} and people who are trained to think in terms of the relativistic interpretation will be quite startled at first. The reason for their surprise is that the FitzGerald-Lorentz contraction is quite real - as real as the contraction of metal rods when their temperature is decreased. Lorentz contraction is a true physical contraction. Within LPI these two examples are not difficult to explain because bodies that move with a velocity relative to the aether are indeed contracted by physical forces. There is a true physical force that \textit{causes} the contraction and it may well break delicate strings and threads. In Minkowksian interpretation the bodies are not three dimensional but four-dimensional objects. And when the objects move it is like seeing them in the four-dimensional space-time from different 'angles'. Thus effects like the above are explained also in Minkowskian interpretation better than the relativistic interpretation. Examples like that show that Minkowskian interpretation has more explanatory power than the relativistic interpretation. And for that reason the practitioners of relativity theory favor the Minkowskian interpretation rather than the relativistic interpretation. 

Therefore these examples show that the relativistic interpretation is explanatorily impoverished as compared with the LPI and the Minkowskian interpretation. However there are more problems. Indeed, since the relativistic interpretation assumes A-theory of time only the present exists. But the very notion of the present (and thus of what exists) is frame dependent. In one reference frame, a person may be shot dead, while in another he may still be alive (not yet shot). If the two reference frames are to have an equal status, then each reference frame is like a new world in which different things are real! Going from one reference frame to another is the same as going from one world to another. Such a pluralistic ontology is fantastic. Even worse, the relativistic interpretation is based upon arbitrary conventions. The relativist believes that he is compelled to choose $\epsilon$ by convention because one cannot establish \textit{empirically} distant simultaneity. However the philosophy behind that is the old defunct philosophy of positivism (according to which things that one cannot measure are meaningless). However, this philosophy has been abandoned\cite{Craig2001} by the majority of the philosophers of science since it is too restrictive and is contrary to the scientific endeavor. A scientist quite often postulates the existence of many things which are not yet empirically established in order to give explanations of a phenomenon - the molecular hypothesis in statistical mechanics has easily explained thermodynamics and chemical reactions well before these molecules were detected directly. Many other examples could be multiplied - the Higgs boson, great many elementary particles, chemical elements, the prediction of the existence of the planet Neptune, etc. In addition, positivism confuses epistemology (what we can know) with ontology (what exists).

Neither does Minkowskian interpretation solves the above problems satisfactorily because it is beset with other difficulties. Indeed, the first difficulty is the union of space with time. Just because one can write space and time coordinates on the same coordinate system, one cannot consider the space-time as real. One can unite pressure and volume on a single coordinate system. This does not mean that there is such a thing as a pressure-volume space. Neither does it help to claim that space-time is different than volume-pressure space by the presence of four-dimensional metric. But how has one detected this metric in the first place? One had to apply the clock synchronization procedure first, which is quite arbitrary and rests on arbitrary conventions (the choice of $\epsilon$) and on defunct positivistic principle. Different conventions of $\epsilon$ will lead to different metrics (Reichenbach\cite{Reichenbach1957} gives such examples). In addition, if one is to accept the realism of the space-time one has to accept the possibility that $ds^{2}<0$, i.e. space-like four-dimensional intervals exist and are complex numbers, which is quite incredible. But even worse than that is the acceptance of B-theory of time which flies in the face of our experience of time. B-theory assumes that past and future exist, that there is a hypothetical possibility of time-travel in the past. But there is no evidence of such things. In fact, one can argue that the A-theory of time is a properly basic belief\cite{Craig2001} and the burden of proof lies upon the shoulders of the B-theorist. What is the evidence for B-theory? There is none. B-theory is simply postulated without any evidence. Thus, it is quite save to say that space-time is merely a good instrument, already used in Newtonian physics and is not to be accepted as the true reality.

Things are aggravated greatly if quantum mechanical considerations are taken into account. Bell's inequalities seem to point that only non-local hidden variable theories are a reasonable alternative to Copenhagen interpretation, while these theories seem to be in great deal of tension with relativity theory. This is not so however in LPI, which can easily accommodate superluminal velocities with Lorentz transformations. Quoting Bell \cite{Bell1986}: "I think it's a deep dilemma, and the resolution of it will not be trivial; it will require a substantial change in the way we look at things. But I would say that the cheapest resolution is something like going back to relativity as it was before Einstein, when people like Lorentz and Poincare thought that there was an aether - a preferred frame of reference - but that our measuring instruments were distorted by motion in such a way that we could not detect motion through the aether...The reason I want to go back to the idea of an aether here is because these EPR experiments there is the suggestion that behind the scenes something is going faster than light. Now, if all Lorentz frames are equivalent, this also means that things can go backward in time...this introduces great problems, paradoxes of causality, and so on. And so it is precisely to avoid these that I want to say there is a real causal sequence which is defined in the aether".

The introduction of general relativity as a proof that the space-time is necessary not just as an instrument but as a reality is also implausible since there is a perfectly reasonable field theoretical explanation of gravity, the so called bimetric theory of gravity \cite{Rosen1940, Cavalleri1980}. Such a bimetric approach to gravity makes possible to consider gravity as a field and energy-momentum tensor can be written. Not only that but the field approach unites all forces of nature under a single unified framework.  Even more, according to Logunov \cite{Logunov1988} one is compelled to consider gravity as a field in flat space-time such that the gauge is organically built into the theory. Otherwise Einstein's gravity equations will not give unique predictions.


\end{document}